\documentclass[journal=apchd5,manuscript=article]{achemso}

\usepackage[version=3]{mhchem} 

\usepackage{graphicx}
\usepackage{dcolumn}
\usepackage{bm}

\author{Kirill Koshelev}
\affiliation{Nonlinear Physics Center, Research School of Physics, Australian National University, Canberra, ACT 2601, Australia}
\alsoaffiliation{equal contribution}
\email{kirill.koshelev@anu.edu.au}

\author{Pavel Tonkaev}
\affiliation{Nonlinear Physics Center, Research School of Physics, Australian National University, Canberra, ACT 2601, Australia}
\alsoaffiliation{equal contribution}

\author{Yuri Kivshar}
\affiliation{Nonlinear Physics Center, Research School of Physics, Australian National University, Canberra, ACT 2601, Australia}
\email{yuri.kivshar@anu.edu.au}

\title{Nonlinear chiral metaphotonics}


\begin{document}

\begin{abstract}
We review the physics and some applications of photonic structures designed for the realisation of strong {\it nonlinear chiroptical response}. We pay much attention to the recent strategy of utilizing different types of {\it optical resonances} in metallic and dielectric subwavelength structures and metasurfaces, including surface plasmon resonances, Mie resonances, lattice guided modes, and bound states in the continuum. We summarize earlier results and discuss more recent developments for achieving large circular dichroism combined with the high efficiency of nonlinear harmonic generation. 
\end{abstract}

\newpage


\section{Introduction} 

Chirality of an object is defined by complete absence of its mirror symmetries. With this property, chiral structures can be divided in two groups of enantiomers representing mirror images of each other. Natural chirality is associated with a microscopic material structure of chiral ions and molecules. The degree of chirality can be probed by circularly polarized light which allows for distinguishing enantiomers via chiroptical effects arising due to different interaction of light with left and right polarizations and media~\cite{barron2009molecular}. In practice, electromagnetic waves with circular polarization are widely employed in a variety of applications in photonics, biology, and chemistry, including ultrafast magnetic imaging~\cite{kfir2017nanoscale}, and spectroscopy~\cite{ranjbar2009circular}.  

Chiroptical effects manifesting optical activity can be expressed in terms of {\it optical rotatory dispersion} (ORD), in which the wave polarization plane is rotated after propagation through a chiral medium, and {\it circular dichroism} (CD)~\cite{berova2000circular}, which describes asymmetric transmission or absorption of circularly polarized light inside a chiral medium. In natural materials, optical activity is very weak.  Over the last two decades, artificial materials with chirality engineered at the macroscopic level were introduced to enhance weak natural chiroptical effects. In earlier studies, three-dimensional chiral metamaterials were shown to enhance ORD and CD in the linear regime, and  provide a simpler route to negative refraction~\cite{wang2009chiral, oh2015chiral}. Due to the complexity of experimental realization of bulk chiral metamaterials, the research focus shifted towards the study of planar chiral structures composed of ordered arrays of subwavelength scatterers, which were demonstrated to change the polarization state of light in a way similar to three-dimensional chiral media~\cite{papakostas2003optical,kuwata2005giant,konishi2008observation, volkov2009optical}. 

Further improvement of efficiency of linear chiroptical response was achieved in the last decade by employing resonant metaphotonic structures made of plasmonic and dielectric materials~\cite{valev2013chirality, hentschel2017chiral}. In particular, very recently planar dielectric metastructures hosting photonic Mie resonances were shown to enhance the local chirality density and provide selective coupling to left-circularly polarized (LCP) and right-circularly polarized (RCP) waves~\cite{gorkunov2020metasurfaces,overvig2021chiral}. This led to far-reaching implications for chiral emission~\cite{zhang2022chiral}, chiral sensing~\cite{mohammadi2018nanophotonic, solomon2018enantiospecific}, characterization of quantum light~\cite{wang2023characterization}, and other areas.

Natural nonlinear optical activity is substantially more pronounced than linear chiroptical effects due to the high sensitivity of optical harmonics generated by chiral light to the molecular and structural asymmetry~\cite{verbiest1999light}. Such high sensitivity attracted special interest to {\it nonlinear chiroptical effects}, such as second- and third-harmonic CD, and led to comprehensive theoretical studies and experimental analysis of nonlinear optical activity for bulk media and planar interfaces in 1990s and early 2000s~\cite{petralli1993circular,byers1994second,maki1995surface}. High increase of nonlinear optical activity was demonstrated for supramolecular arrays in helicene films of chloroform~\cite{verbiest1998strong}. Similar to linear effects, the enhancement of nonlinear chiroptical effects beyond the limit of natural material response can be achieved by engineering chirality at macroscopic scale~\cite{haupert2009chirality, li2017nonlinearR, konishi2020tunable, rodrigues2022review}. What different to the linear regime is that strong nonlinear optical activity also requires the high efficiency of nonlinear conversion. For this reason, resonances of metaphotonic structures are expected to play a crucial role as they were shown to improve the frequency conversion efficiency dramatically by virtue of electric and magnetic hot-spots enhancing the light-matter interaction~\cite{luk2010fano,koshelev2020dielectric}. 

In this Perspective, we overview the development and major advances in the field of nonlinear chiral metaphotonics focusing on resonances of light and matter and their effect on nonlinear optical activity in nanostructures and metasurfaces. We start from a brief description of different types of resonances that can be realized in meta-optics with plasmonic and dielectric chiral metastructures, varying from surface plasmon polaritons to optical bound states in the continuum. Next, we describe briefly the history of research on nonlinear chiral photonics starting from non-resonant planar plasmonic arrays with macroscopic chiral geometry and moving to resonant metasurfaces which shape the nonlinear chiroptical response by virtue of plasmonic resonances. We then discuss how the efficiency of nonlinear chiroptical response can be improved by employing dielectric Mie-resonant metasurfaces and nanoantennas, which sustain substantially higher damage thresholds and provide additional degrees of freedom via excitation of optically-induced magnetic resonances. We discuss how collective resonances of dielectric metasurfaces with a large quality factor can be employed to achieve maximal nonlinear chirality. Throughout the paper, we deliberate the role of the microscopic and macroscopic symmetries, geometrical phases, and various resonant effects. Finally, we conclude with an outlook of several novel opportunities in this rapidly expanding research area. 



\section{Resonances in metaphotonic structures}

\begin{figure}[t]
   \centering
    \includegraphics[width=0.95\linewidth]{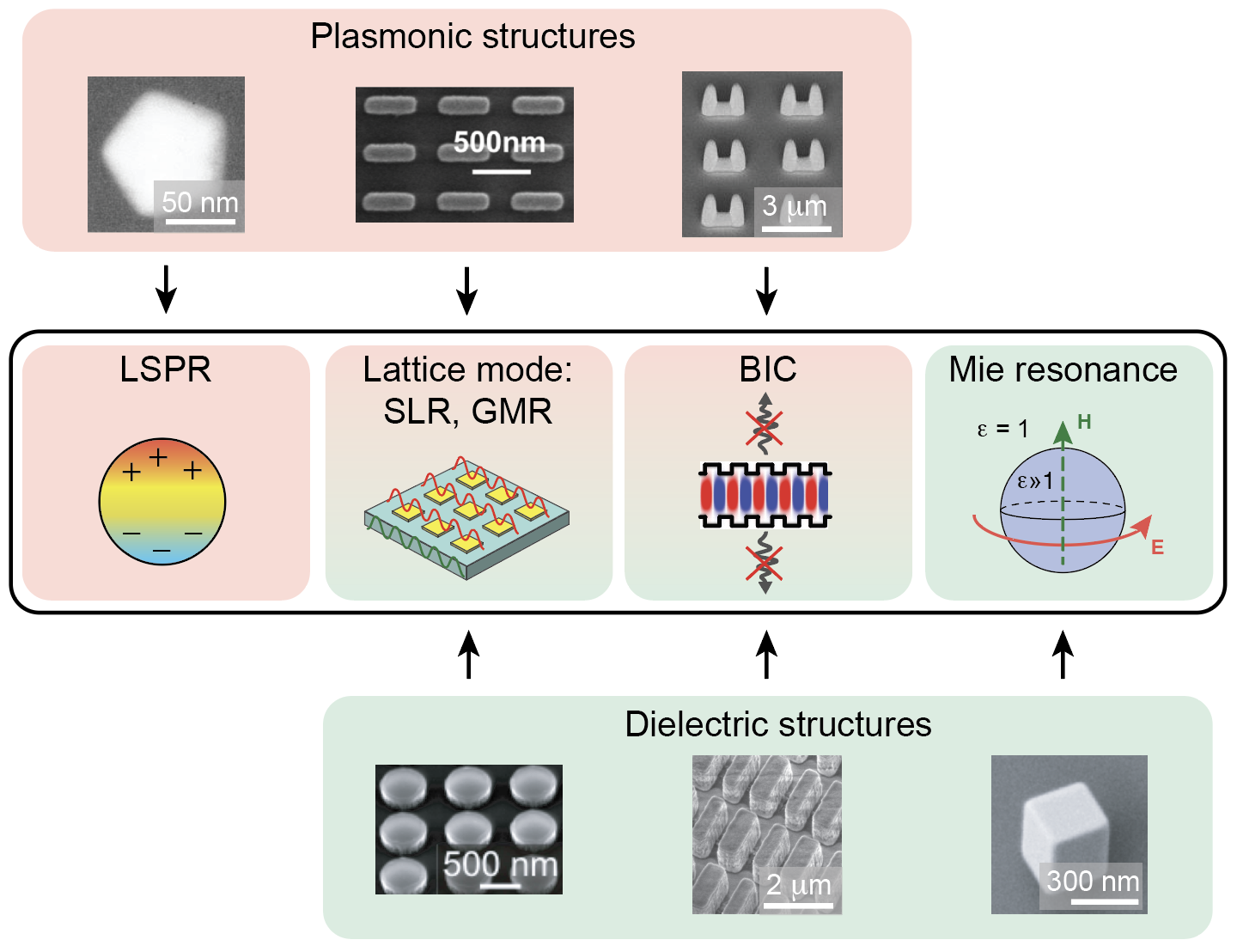}
    \caption{{\bf Resonances in metaphotonic structures.} Top (red background): SEM images of selected plasmonic metastructures (red) with LSPR, SLR, and BIC resonances, respectively. Middle: types of resonances in plasmonic (red), dielectric (green) and both types of material (red-green). These include LSPR, lattice mode (SLR and GMR), BIC, and Mie resonance. Bottom (green background): SEM images of selected dielectric metastructures  with GMR, BIC, and Mie resonances. SEM images are adapted from Refs.~\cite{mock2002shape, giannini2010lighting, liang2020bound, staude2013tailoring, zograf2022high, tiguntseva2020room}.}
    \label{fig1}
\end{figure}

Chiroptical properties of metaphotonic structures composed of isolated nanoparticles or their arrays in the form of photonic crystals and metasurfaces are defined by the meta-atom geometry, lattice arrangement, and material properties. Due to the subwavelength size of individual scatterers, the resonant wavelengths in materials are comparable with the meta-atom characteristic size, allowing to control resonance properties via geometry. Here, we briefly overview the most typical resonances in plasmonic and dielectric metastructures, shown schematically in the central panel of Fig.~\ref{fig1} (with the images adopted from several papers~\cite{mock2002shape, giannini2010lighting, liang2020bound, staude2013tailoring, zograf2022high, tiguntseva2020room})  to better outline the differences they can provide in terms of nonlinear optical activity.

The fundamental modes of subwavelength plasmonic nanostructures are associated with localized surface plasmon resonances (LSPRs) representing nonpropagating coherent excitations of the free electron density coupled to the radiation continuum~\cite{maier2007plasmonics}. The LSPR frequency depends on the plasma frequency of the  material, shape of a nanoparticle, and its size. For small particles, the LSPR lifetime is determined by the nonradiative decay due to the material absorption. With an increase of the volume, the radiative loss channel starts to dominate the mode lifetime. The typical quality factor (Q factor) of dipolar and quadrupolar LSPRs is in the range $2-10$. Plasmonic resonances in isolated nanoparticles could provide the local electric field enhancement up to several hundred due to a deeply subwavelength mode volume. An example of the scanning electron micrograph (SEM) image of silver nanoparticle with a $100$~nm lateral size hosting LSPR at $520$~nm~\cite{mock2002shape} is shown on the left in the top panel of Fig.~\ref{fig1}.

Ordered arrays of plasmonic nanoparticles can host collective lattice modes, which can be generally divided into {\it two classes}: localized surface lattice resonances (SLRs), also called {\it lattice surface modes}, and nonlocal {\it guided-mode resonances} (GMRs). The SLRs originate from the near- and far-field coupling of LSPs of individual meta-atoms via diffractive orders of the lattice~\cite{kravets2018plasmonic}. Formation of SLRs leads to the suppression of radiative losses of single LSPRs, and thus results in a relatively large $Q$ factor of the lattice mode up to several hundred units. The resonant wavelength of SLRs depends on the lattice parameters including period and mutual arrangement, as well as meta-atom material and geometry. The electromagnetic fields of SLRs are predominantly localized at the nanoparticle surfaces. An example of the SEM image of the plasmonic metasurface composed of a rectangular array of gold nanobars~\cite{giannini2010lighting} and supporting SLR at about $900$~nm with $Q=11$ is shown in the middle of the top panel of Fig.~\ref{fig1}. The GMR-type lattice modes are formed when the plasmonic array is positioned on top a waveguiding dielectric layer, and the diffractive coupling between individual LSPRs occurs via the waveguide modes of the slab.  For GMRs, the near-fields are localized both on the particle surface and inside the waveguiding layer. 

{\it Plasmonic bound states in the continuum} (BICs) represent a special class of plasmonic lattice modes with the Q factor which tends to infinity due to destructive interference of radiation of the constituent LSPRs in the far-field. The BICs are associated with non-local bound lattice modes with frequencies lying in the radiation continuum~\cite{koshelev2020dielectric}. The formation of quasi-BICs with a finite yet large Q factor can be achieved by introducing an asymmetry of meta-atoms or extrinsic asymmetry via the experimental setup. The first observation of plasmonic BICs was reported only very recently~\cite{liang2020bound}, the SEM image of the corresponding plasmonic metasurface composed of a rectangular array of U-shaped particles is shown on the right of the top panel of Fig.~\ref{fig1}. The metasurface supports a high-$Q$ plasmonic quasi-BIC in the mid-IR frequency range. Recently, it was shown~\cite{bin2021ultra} that plasmonic BICs are related to out-of-plane SLRs, and their Q factor can exceed $2000$. 

Dielectric resonant structures emerged recently as a new platform in metaphotonics that allows effective localization of light in a bulk of  material and additional degrees of freedom through the generation of artificial magnetic response~\cite{koshelev2020dielectric}. The main building block of dielectric metastructures is an isolated dielectric nanoparticle hosting geometric Mie resonances, also called {\it Mie modes}. The Mie resonance wavelengths are defined as fractions of the lateral size of the nanoparticle multiplied by its refractive index. For high-index nanoresonators, the Mie modes with large values of the $Q$ factor of up to several dozen can be readily realized in the visible and near-infrared (near-IR) frequency ranges. Compared to metals~\cite{kern2015limitations}, high-index dielectric materials are free of material losses in the most relevant frequency ranges, which increases their potential for chiral nonlinear metaphotonics due to high damage thresholds. An example of SEM image of a high-index dielectric nanocube composed of a lead-halide perovskite~\cite{tiguntseva2020room} supporting a high-order Mie mode at $535$~nm with $Q=15$ is shown on the right of the bottom panel of Fig.~\ref{fig1}.

Dielectric meta-atoms arranged in ordered arrays provide an additional way of control of chiroptical response via collective lattice modes. Similar to plasmonic lattices, dielectric metasurfaces, photonic crystal slabs and membranes can host highly-localized lattice Mie modes with weak near-field coupling and highly nonlocal GMRs with a large $Q$ factor produced by strong diffractive coupling in the near- and far-field. In the case of GMRs, wave guiding can be realized not in the substrate but in the periodic array itself via collective oscillations of the polarization density. An example of a SEM image of a high-index dielectric metasurface composed of a square lattice of Si disks embedded in a low-index medium is shown on the left of the bottom panel of Fig.~\ref{fig1}. The metasurface supports almost degenerate magnetic and electric Mie lattice modes with the Q of order of $10$ in the near-IR wavelength range.

The lattice modes of dielectric arrays can transform to high-Q photonic BICs and quasi-BICs under specific conditions, including unit cell asymmetry, parametric tuning, or mode strong coupling~\cite{koshelev2018asymmetric,koshelev2020dielectric}. The $Q$ factor of the quasi-BIC mode can exceed $10^4$ in broken-symmetry dielectric metasurfaces~\cite{liu2019high}, and $10^5$ in suspended membranes~\cite{jin2019topologically}. Chiral BICs can be realised in dielectric metasurfaces intrinsically, by making three-dimensional chiral meta-atoms~\cite{gorkunov2020metasurfaces,overvig2021chiral,zhang2022chiral,chen2023observation}, or extrinsically, in planar metasurfaces with achiral meta-atoms with broken in-plane mirror symmetries placed on a substrate~\cite{shi2022planar}. In addition to a high $Q$ factor, BICs allow for advanced polarization control provided they form topological vortexes in the reciprocal space. An example of a SEM image of a high-index dielectric metasurface composed of a square lattice of Si bar dimers with different widths is shown in the middle of the bottom panel of Fig.~\ref{fig1}. The metasurface supports a quasi-BIC with $Q=200$ in the mid-IR wavelength range.

In scattering, LSPR, Mie, lattice and quasi-BIC resonances manifest normally themselves with a pronounced feature with an asymmetric Fano lineshape due to interference with broad non-resonant modes, and often termed generically as Fano resonances~\cite{limonov2017fano}. In particular, Fano resonances originating from quasi-BIC modes are sometimes referred to as high-Q or sharp Fano resonances~\cite{luk2010fano}. Moreover, it can be shown that for quasi-BICs the Fano asymmetry parameter is connected to the Q factor and diverges when the mode reaches its largest Q factor due to the mode strong coupling~\cite{melik2021fano}. Also, quasi-BICs have been shown to govern the related effects of metamaterial-induced transparency and plasmonic-induced transparency~\cite{koshelev2018asymmetric}. 


\section{Nonlinear chiral plasmonic metasurfaces}

\subsection{Non-resonant metallic metasurfaces}

\begin{figure}[t]
   \centering
    \includegraphics[width=0.95\linewidth]{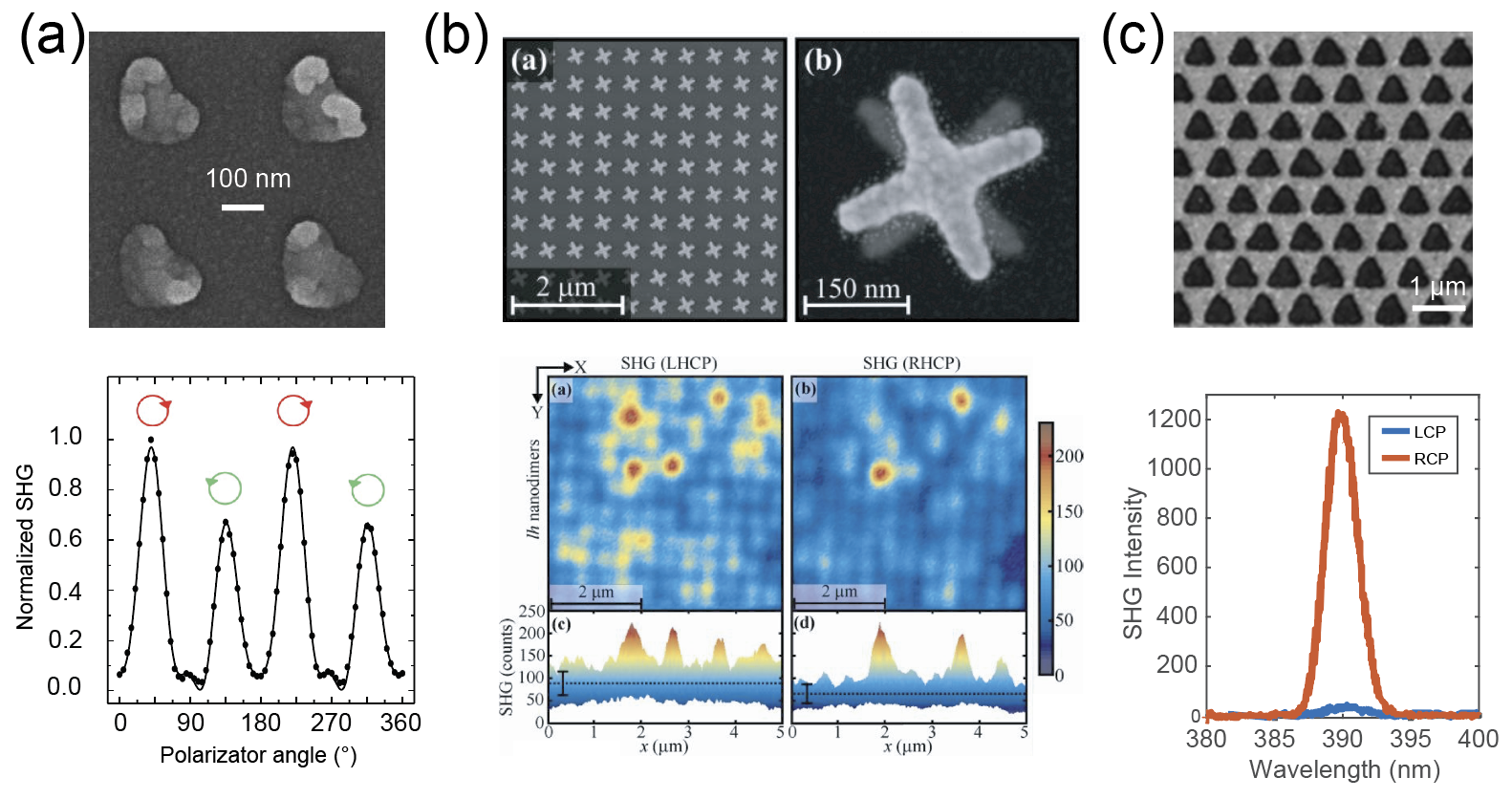}
    \caption{{\bf Nonlinear chiroptical response from nonresonant metallic metastructures}. (a) Metasurface composed of a square array of L-shaped gold nanoparticles with small defects~\cite{canfield2006macroscopic}. Top: SEM image; bottom: non-resonant SH signal spectrum dependence on the orientation angle of the input beam polarizer. (b) Metasurface composed of gold twisted-cross nanodimers with three-dimensional chirality\cite{huttunen2011nonlinear}. Top: SEM image; bottom: non-resonant SH microscopy maps for left-handed nanodimers for LCP and RCP pump, respectively. (c) Metasurface composed of triangular nanoholes in gold arranged in a honeycomb lattice. Top: SEM image; bottom: polarization-resolved SH spectra for the LCP excitation in the near-IR range~\cite{konishi2014polarization}.}
    \label{fig2}
\end{figure}

Next, we discuss the historical development of studies on nonlinear chiroptical metastructures outlining several primary approaches allowing to achieve and control nonlinear optical activity. One of the first studies of chiral nonlinear response in nanostructured arrays were done in early 2000s for non-resonant metallic metasurfaces. Limited fabrication quality of metallic nanostructures resulted in small nanoscale defects which provided a natural way of achieving three-dimensional nonlinear optical activity~\cite{canfield2006chirality}. The overall second-order response was shown to be related to a complicated interplay between plasmonic resonances of the particles and their structural defects. One of the pioneering works in this field~\cite{canfield2006macroscopic} was devoted to the study of the second-order nonlinear characteristics of an array of L-shaped gold nanoparticles, shown in the SEM image in the upper panel of Fig.~\ref{fig2}a. The structure supported fundamental plasmonic resonances at $1020$~nm for one linear polarization, and at $640$~nm and $800$~nm for the other polarization. Despite the presence of geometrical dichroism and resonances, the nonlinear response and activity were produced predominantly at the defects. The bottom panel of Fig.~\ref{fig2}a shows the normalized second-harmonic (SH) signal for different orientations of the quarter-wave plate modifying the polarization of the incident beam from RCP to LCP and back. The signal was collected in the nonresonant regime at $1060$~nm. The resulting SH efficiency was different for both circular polarizations indicating a nonlinear SH-CD. The defect contributions to the nonlinear optical response in such metasurfaces were later explained by effective higher multipolar terms in the nonlinear response~\cite{kujala2007multipole}. This resulted in consequent studies on the separation of magnetic dipolar and electric quadrupolar effective nonlinear responses~\cite{kujala2008multipolar,zdanowicz2011effective}. Finally, with the improvement of the sample quality, the dipole limit of the effective nonlinear chiroptical response was reached~\cite{czaplicki2011dipole}.

Based on these developments, one of the key experimental methods to probe the nonlinear chirality of individual nanostructures with the SHG microscopy was applied in a series of studies. In particular, in paper \cite{huttunen2011nonlinear} the authors studied the nonlinear chiroptical response of metallic double-layer metasurfaces with a three-dimensional chirality composed of gold twisted-cross nanodimers. The corresponding SEM images are shown in the top panel of Fig~\ref{fig2}b. The meta-atoms can be left- or right-handed depending on the mutual orientation of the crosses. The bottom panel of Fig~\ref{fig2}b shows the nonresonant SH microscopy images of the left-handed nanodimers for the LCP and RCP pump, respectively, at the pump wavelength of $1060$~nm. The metasurface demonstrated the enhancement of second-harmonic generation (SHG) when the polarization of incident light coincided with the handedness of the nanostructure.


Early works in the field of nonlinear chirality in natural optical active crystals showed that the symmetry of the molecules defines the nonlinear optical activity via series of selection rules~\cite{simon1968second}. For bulk media it was shown that harmonic generation for light with circular polarization follows the selection rule $Np_0=ms+p_N$, where $N$ is the order of harmonic generation, $m$ is the order of rotational symmetry of the molecules, $p_0=\pm1$ and $p_N=\pm1$ are the incident and harmonic wave handedness, respectively, and $s=0,\pm 1,\dots$ is an arbitrary integer~\cite{alon1998selection,higuchi2011selection}.


In early 2010s, these selection rules were generalized for macroscopic symmetry of metastructures outlining a key approach to control the chirality in metaphotonics. In particular, it was predicted that the selection rules can be also applied for harmonic generation in  plasmonic metasurfaces composed of ordered meta-atom arrays with specific lattice symmetry. For metallic non-resonant metasurfaces this effect was first shown in a triangle hole-arrayed gold nanostructure~\cite{konishi2014polarization}, which SEM image is shown in the top panel of Fig.~\ref{fig2}c. The selection rules for $N=2$, $m=3$ can be fulfilled only for $s=1$ and $p_0=-p_N$, which means that for a circularly polarized pump only a counter-polarized SH signal can be generated. The measured SH spectra of LCP and RCP components from this metasurface under the RCP excitation at $780$~nm are shown in the bottom panel of Fig.~\ref{fig2}c. The nonlinear transmission measurements showed that the RCP pump generates a strong SH LCP signal with a very weak RCP component, which confirms the selection rules. This effect is also preserved for the opposite case when the excitation beam is LCP. The weak non-zero co-polarized SH signal component arose only due to distortion of the shape from triangular geometry. For comparison, cases of L-shaped and circular holes in a triangular lattice were studied. For L-shaped structures, the co-polarized SH component was as strong as the cross-polarized, while for circular holes the SH was zero, as it was prohibited by selection rules due to six-fold rotational symmetry.




\subsection{Resonant plasmonic metasurfaces}

\begin{figure}[t]
   \centering
    \includegraphics[width=0.95\linewidth]{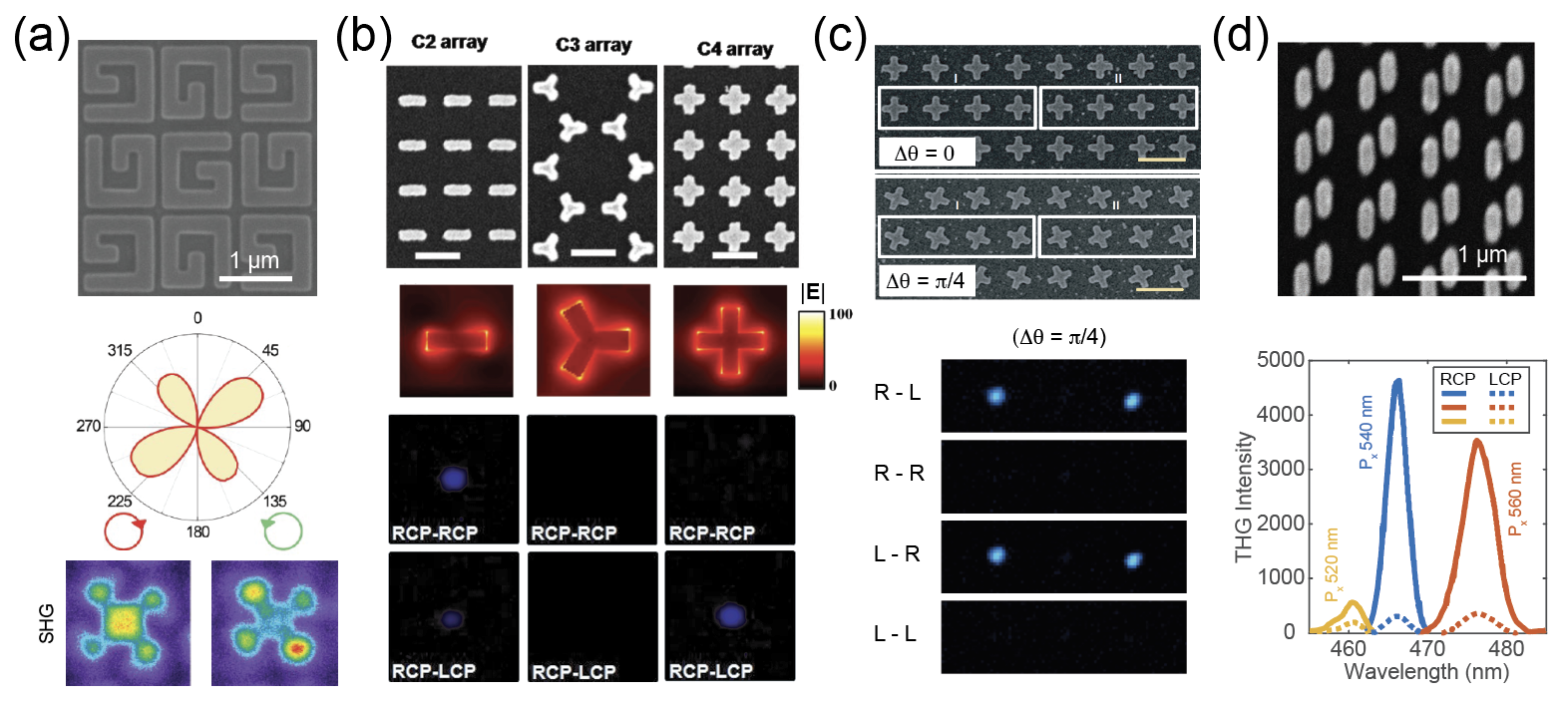}
    \caption{{\bf Nonlinear chiroptical response from resonant plasmonic metasurfaces}. (a) Plasmonic metasurface with LSPR resonances composed of super-cells pf gold G-shaped meta-atoms arranged in a square lattice with $C_4$ rotational symmetry~\cite{valev2009plasmonic}. Top: SEM image; bottom: SH signal dependence on the pump polarization and SH microscopy maps for LCP and RCP excitation. (b) Nonlinear metasurfaces supporting plasmonic lattice modes with engineered macroscopic $C_2$,$C_3$ and $C_4$ rotational symmetry~\cite{chen2014symmetry}. Top: SEM image; center: mode field profiles; bottom: TH signal maps for LCP and RCP excitation for corresponding metasurfaces. (c) Nonlinear Pancharatnam-Berry phase metasurface composed of gold crosses hosting LSPR resonances in the near-IR range~\cite{li2015continuous}. Top: SEM image; bottom: experimental diffraction patterns of TH signals for the metasurfaces with geometric phase of $90$ degrees. (d) Integrated nonlinear metasurface supporting a hybrid plasmonic-photonic modes. Top: SEM image; bottom: TGH intensity for LCP and RCP excitation depending on the period of plasmonic metasurface $P_x$~\cite{wang2019observation}.}
    \label{fig3}
\end{figure}

One of the first works demonstrating nonlinear CD in resonant plasmonic metasurfaces was done in 2009 for a metasurface composed of gold G-shaped meta-atoms~\cite{valev2009plasmonic}, which SEM image is shown in the top panel of Fig.~\ref{fig3}a. The meta-atoms were arranged in a square lattice, with a super-cell of four elements rotated by $90$ degrees with respect to the neighboring atom. By this, the metasurface designed to possess a macroscopic rotational symmetry $C_4$ induced at the super-cell level. The structures were pumped at $800$ nm close to the LSPR resonance frequency. Change of the pump beam polarization with a quarter-wave plate caused the change in second harmonic intensity, as shown in the bottom panel of Fig.~\ref{fig3}a, which resulted in the resonant SH-CD. The SHG microscopy images showed that the SH signal originated from the field hotspots attributed to plasmonic modes. This was the first work demonstrating that the arrangement of the nanostructures plays a crucial role since upon reordering them the SH-CD effect vanishes. Depending on the handedness of the nanostructures, the SHG sources exhibited a ratchet wheel pattern with a different sense of rotation. This allows for the possibility to optically determine the handedness of a material by observing it with one circularly polarized incident beam.

Later, it was shown that even pumped by linearly polarised light the gold specially oriented G-shaped nanostructures demonstrate SH hotspots, whose arrangement is dependent on the handedness~\cite{valev2010asymmetric}. Another proposed method of chirality determination by linear light was based on SHG depending on the rotation of the sample~\cite{valev2010asymmetric}. This was the first observation of a nonlinear chiroptical effect with magnetic dipoles.  Later, it was shown that superchiral light, which is the local enhancement of the chirality density above the value for circularly polarized light in vacuum, and the SHG process are fundamentally linked via their dependence on the electric quadrupolar response of the resonant structures~\cite{valev2014nonlinear}. Additionally, the study of CD in an optical SHG in planar symmetrical arrays of gold G-shaped nanostructures showed that nonlinear CD demonstrated a strong dependence of the value and the sign of the effect on the angle of incidence of the incident wave~\cite{mamonov2014anisotropy}. 




The macroscopic selection rules for nonlinear chiroptical response of resonant plasmonic metasurfaces were first outlined in the work~\cite{chen2014symmetry}. The SEM images of gold metasurfaces studied in the paper are shown in the top panel of Fig.~\ref{fig3}b. The geometry of meta-atoms and their arrangement were chosen to reproduce the rotational symmetry of the second, third, and fourth order, respectively. The structures supported broad fundamental LSPRs at about $1300$~nm, with the field profiles shown in the central panel of Fig.~\ref{fig3}b. The bottom panel of Fig.~\ref{fig3}b shows experimentally measured THG images for different circular polarizations of pump and filtered harmonic polarization. The measurements confirm that selection rules applied to the lattice symmetry of the metasurface, that state that RCP or LCP harmonics can be generated only in two- and fourfold rotationally symmetric structures, but the chiral THG is forbidden for the threefold rotational symmetry configuration. It was also shown that this effect can be weakly disturbed at the microscopic level by the resonant contribution of modes, which near-field has out-of-plane non-circular polarization components. A separate study was devoted to the nonlinear chiroptical response of quasi-crystal resonant plasmonic metasurfaces. In the paper~\cite{tang2019quasicrystal} gold meta-atoms in nonlinear chiral metasurface arranged in a Penrose quasi-crystal were shown to demonstrate chiral SHG, which inverts circularly polarised pump from RCP to LCP and vice versa, similar to selection rules for regular photonic lattices.

One of the pioneering works demonstrating giant nonlinear optical activity in a plasmonic metasurface was done in 2012 for a metasurface composed of a square lattice of split-ring hole resonators~\cite{ren2012giant}. The demonstrated resonant nonlinear optical activity was caused by direct two-photon absorption. This effect was shown to manifest itself via power-dependent circular birefringence and circular dichroism. The chirality was induced by breaking the symmetry extrinsically with an oblique incidence of the incident beam. A related observation of nonlinear SHG from achiral metasurfaces with broken in-plane symmetries was done later in 2016~\cite{chen2016giant}. It was shown that intrinsic or extrinsic geometrical three-dimensional chirality is not necessary for introducing strong circular dichroism for harmonic generations. Specifically, a near-unity circular dichroism for both SHG and THG attributed to nonlinear symmetry breaking  was demonstrated for engineered resonant plasmonic metasurfaces with broken in-plane mirror symmetries.

Another prominent concept in nonlinear chiral metaphotonics is Pancharatnam-Berry phase applied for control of harmonic signal polarisation and direction pumped by circular light~\cite{tymchenko2015gradient}. The local effective nonlinear dipole moment $\mathbf{p}_m$ for a LCP or RCP pump incident along the rotational axis of a meta-atom, can be expressed in the basis of circularly polarized waves as $\mathbf{p}_m(\varphi)=\alpha^{(m)}(\varphi)\mathbf{E}^m$. Here, $\varphi$ is the angle of meta-atom rotation with respect to the primary lattice axes, $\alpha^{(m)}$ is polarizability constant for the $m$-th order nonlinear process, and $\mathbf{E}$ is the pump electric field with LCP or RCP polarization. The polarizability can be expressed as $\alpha^{(m)}(\varphi)\propto \exp{[i(m\pm1)\varphi]}$ for the harmonic photons with cross- and co-polarization of the electric field, respectively. Thus, by controlling the geometrical orientation it is possible to control the phase of the nonlinear signal and construct functional metasurfaces. One of the pioneering experimental demonstrations of the Pancharatnam-Berry metasurfaces was done for gold plasmonic metasurface for continuous control of the nonlinearity phase for harmonic generations~\cite{li2015continuous}. The SEM images of the studied structures are shown in the top panel of Fig.~\ref{fig3}c. The structures supported broad dipolar LSPR modes at about $1220$~nm. Measured diffraction patterns of TH signals shown in the bottom panel of Fig.~\ref{fig3}c for the metasurface with meta-atoms rotated by $90$ degrees demonstrated selective generation of $\pm1$ diffraction orders. Later, using a similar concept, simultaneous control of spin and orbital angular momentum of SH signal in plasmonic metasurface was shown~\cite{li2017nonlinear}. Later, this physics was generalized for non-dipolar responses of resonant phase metasurfaces~\cite{gennaro2019nonlinear}. More examples of Pancharatnam-Berry are discussed in the topical review~\cite{li2017nonlinearR}.

One of the essential drawbacks of plasmonic metasurfaces is their two-dimensional nature which prevents them from exhibiting strong nonlinear optical activity. Recently, it was suggested to employ integrated plasmonic-dielectric structures supporting hybrid plasmonic-photonic modes~\cite{wang2019observation}. The authors studied a plasmonic metasurface composed of a square lattice of gold nanodimers on top of a silicon-on-insulator wafer forming a waveguide, the SEM image is shown in the top panel of Fig.~\ref{fig3}d. The parameters of plasmonic lattice resonance were tuned by adjusting the metasurface geometry to couple it with the Fabry-Perot resonance of the waveguiding Si layer, which exhibited a broad dip in reflection at $1410$~nm. The resulting hybrid mode exhibited a Q factor of about $100$. The measured TH signal exhibited strong nonlinear CD, which also depended on the periodicity of the plasmonic metasurface, as shown in the bottom panel of Fig.~\ref{fig3}d. Another demonstration of a chiral nonlinear hybrid metasurface was done very recently for Au-ZnO integrated nanostructure~\cite{hong2022chiral}. The structure was composed of nanoholes in gold arranged into a metasurface and covered with ZnO dielectric layer. The hybrid metasurface demonstrated the enhancement of THG efficiency compared to a bare ZnO film and the possibility of multiplexed nonlinear holograms with four channels, which is based on the circular polarisation of the fundamental excitation and hologram signal. Finally, we also mention another example of hybrid metasurfaces called polaritonic metasurfaces. The very recently demonstrated on a platform of resonant plasmonic metasurface on top of a multiple semiconductor quantum well structure demonstrating resonant enhancement of second- and third-order susceptibilities due to intersubband transitions~\cite{kim2020giant}. The structure was pumped in mid-IR range at about $10$~microns wavelength and a combination of the strong nonlinear response of TH and TH signals and high nonlinear CD reaching values up to $1$ was demonstrated.







\section{Resonant dielectric metastructures}

The plasmonic platform has several fundamental limitations hindering strong nonlinear chiroptical effects. The major ones are low damage threshold which is essential for high-intensity laser powers needed for nonlinear operation, and two-dimensional geometry, which inherently prevents out-of-plane symmetry breaking required for strong optical activity. Hybrid platforms discussed in the previous section offer an intermediate solution, however, they are limited due to large footprint or similar problems with heating loss. Dielectric metasurfaces and nanostructures offer a favourable alternative to plasmonic nonlinear metasurfaces because of their robustness to absorption losses, and property to host geometrical resonances of a higher-order that exploit the out-of-plane dimension thus providing strong intrinsic linear and nonlinear chirality. In this section, we overview the recent progress in employing dielectric metastructures for enhancement and control of the nonlinear chiroptical response. The field is very young and only a few demonstrations was done so far. We start from dielectric periodic membranes and oligomers of individual nanoresonators hosting GMR and Mie modes, respectively. 

\begin{figure}[H]
   \centering
    \includegraphics[width=0.95\linewidth]{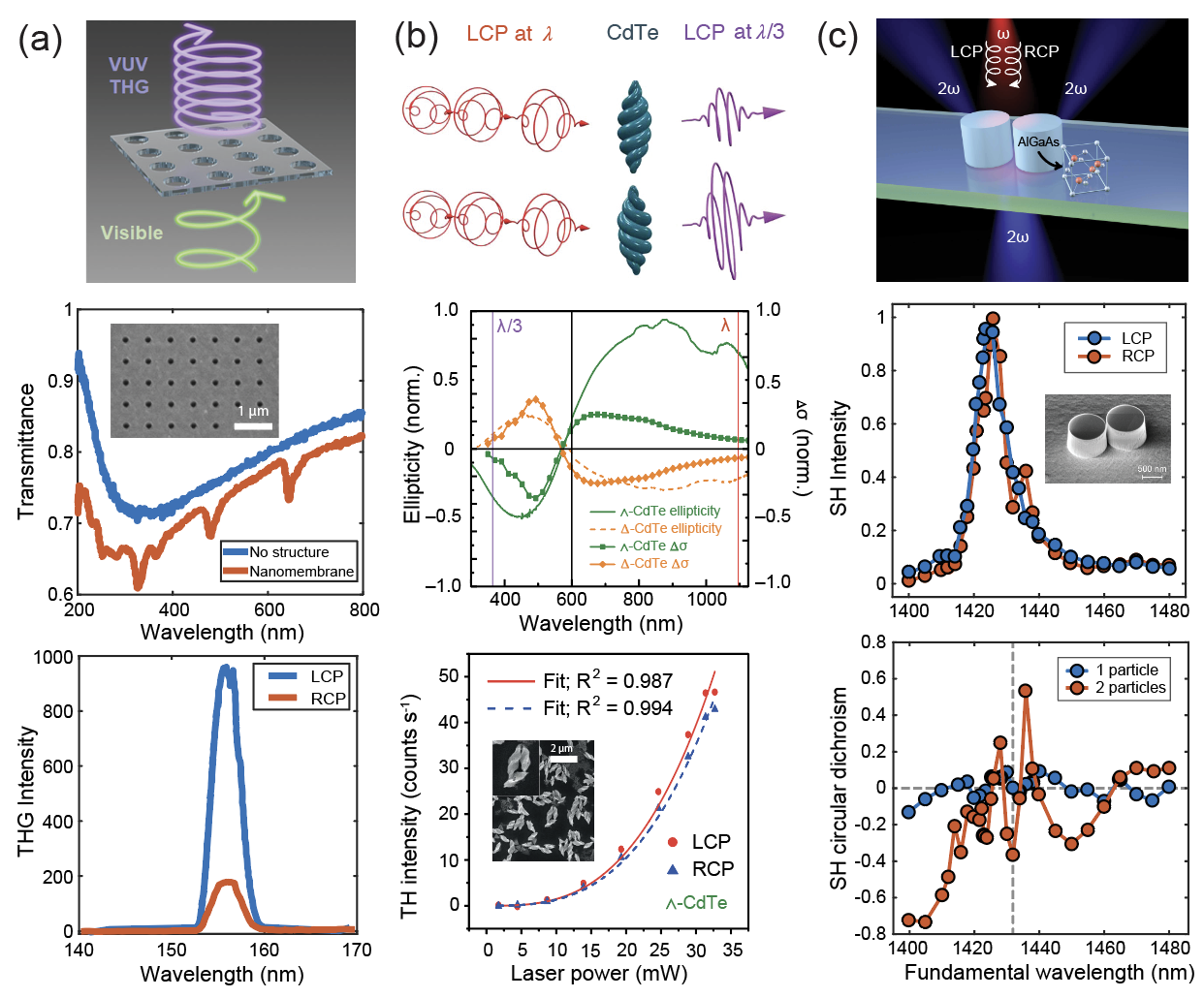}
    \caption{{\bf Nonlinear chiroptical response from resonant dielectric metastructures.} (a) Vacuum UV THG from dielectric $\gamma$-Al$_2$O$_3$ photonic crystal membrane hosting GMR modes~\cite{konishi2020circularly}. Top: schematic; center: linear transmission for structure and unstructured film and SEM image; bottom: LCP and RCP components of the TH intensity for the RCP pump beam. (b) TH CD from Mie-resonant CdTe nanostructured helices~\cite{ohnoutek2022third}. Top: schematic; center: measured linear ellipticity and calculated differential extinction cross-section spectra; bottom: TH intensity in the forward direction vs. pump laser power for the LCP and RCP pump and SEM image. (c) Nonlinear circular dichroism in Mie-resonant dimer by breaking symmetry~\cite{frizyuk2021nonlinear}. Top: schematic; center and bottom: measured SH intensity and SH CD for LCP and RCP pump.}
    \label{fig4}
\end{figure}

The concept of engineering the nonlinear response via the macroscopic symmetry of the photonic structure, that we discussed for plasmonic metasurfaces, was recently transferred to dielectric periodic membranes~\cite{konishi2020circularly}. The authors studied ultraviolet (UV) TH signal from a suspended dielectric $\gamma$-Al$_2$O$_3$ photonic crystal slab with circular air holes arranged in a square lattice so the structure formed a free-standing membrane; the schematic and SEM image are shown in the top and central panel of Fig.~\ref{fig4}a. The structure supported GMR modes in the visible range manifesting themselves as dips in the transmission spectrum, see the central panel of Fig.~\ref{fig4}a. The measured UV TH spectrum for the RCP excitation, shown in the bottom panel of Fig.~\ref{fig4}a, demonstrates the selection rule prohibiting co-polarized RCP THG for the four-fold lattice symmetry $C_4$. The nonzero RCP TH signal is attributed to deviations in the lattice shape transforming it to rectangular geometry. The measured UV TH spectrum for the LCP excitation showed the opposite polarization of the harmonic signal with the same magnitude, confirming that the nonlinear TH CD is zero due to the in-plane mirror symmetries of the unit cell.

Recently, the first observation of the nonlinear optical activity of the SH hyper-Rayleigh scattering, that was predicted more than 40 years ago, was achieved with metallic helical structures suspended in solution~\cite{collins2019first}. Later it was followed by consequent observation of the TH Rayleigh scattering for metallic helices~\cite{ohnoutek2021optical}. Very recently, this concept was transferred to dielectric nanoparticles supporting vectorial Mie resonances, demonstrating strong TH CD in three-dimensional chiral CdTe nanohelices~\cite{ohnoutek2022third}, with schematic shown in the top panel of Fig.~\ref{fig4}b. The left- and right-handed nanohelices were fabricated by variation of the chemical synthesis method and suspended in solution. The structures showed noticeable ellipticity in the linear scattering, which is in agreement with the calculated differential cross sections, as shown in the central panel of Fig.~\ref{fig4}b.  The nanohelices generated the TH preserving the polarization of the circular fundamental beam regardless of the handedness. The magnitude of measured THG was different for the LCP and RCP excitation, depending on the input laser power, see the bottom panel of Fig.~\ref{fig4}b. 

Very recently, the influence of relative atomic lattice orientation and macroscopic meta-molecule geometry on the nonlinear optical activity was studied experimentally. In the work~\cite{frizyuk2021nonlinear} it was shown that an AlGaAs nanodimer, see the top panel of Fig.~\ref{fig4}c, hosting high-Q Mie resonances can provide strong SH CD for specific orientations of the particles with respect to each other. The SH CD was close to zero for a single particle but increased substantially for the dimer with specific relative crystal axes orientation, as shown in the central and bottom panels of Fig.~\ref{fig4}c. This work led to theoretical analysis of conditions for nonzero SH CD in individual nanoparticles and oligomers depending on the atomic lattice orientation and macroscopic symmetry of the meta-molecule~\cite{nikitina2023nonlinear}, based on earlier knowledge of selection rulers for nonlinear processes in nanostructures of different shape~\cite{finazzi2007selection,frizyuk2019second}.

Recent works also include theoretical analysis of intensity-dependent nonlinear chiroptical response driven by GMR modes in Si metasurfaces~\cite{kang2023nonlinear} and numerical study of chiral SHG in LN metasurfaces with Z-shaped meta-atoms in a square lattice on gold substrates~\cite{kim2020dielectric}. 


\section{Metasurfaces with bound states in the continuum}

One of the most promising concepts for chiral nonlinear metaphotonics is associated with chiral quasi-BICs in dielectric resonant metasurfaces~\cite{gorkunov2020metasurfaces,overvig2021chiral}. As outlined above, quasi-BICs were experimentally demonstrated to provide very strong field enhancement in the volume of nanostructures leading to the effective generation of optical harmonics, self-action effects, and high-harmonic generation~\cite{koshelev2019nonlinear, sinev2021observation, zograf2022high}. The combination of chiral-BICs with the enhancement of nonlinear response provides a promising route for giant nonlinear optical activity.

\begin{figure}[t]
   \centering
    \includegraphics[width=0.95\linewidth]{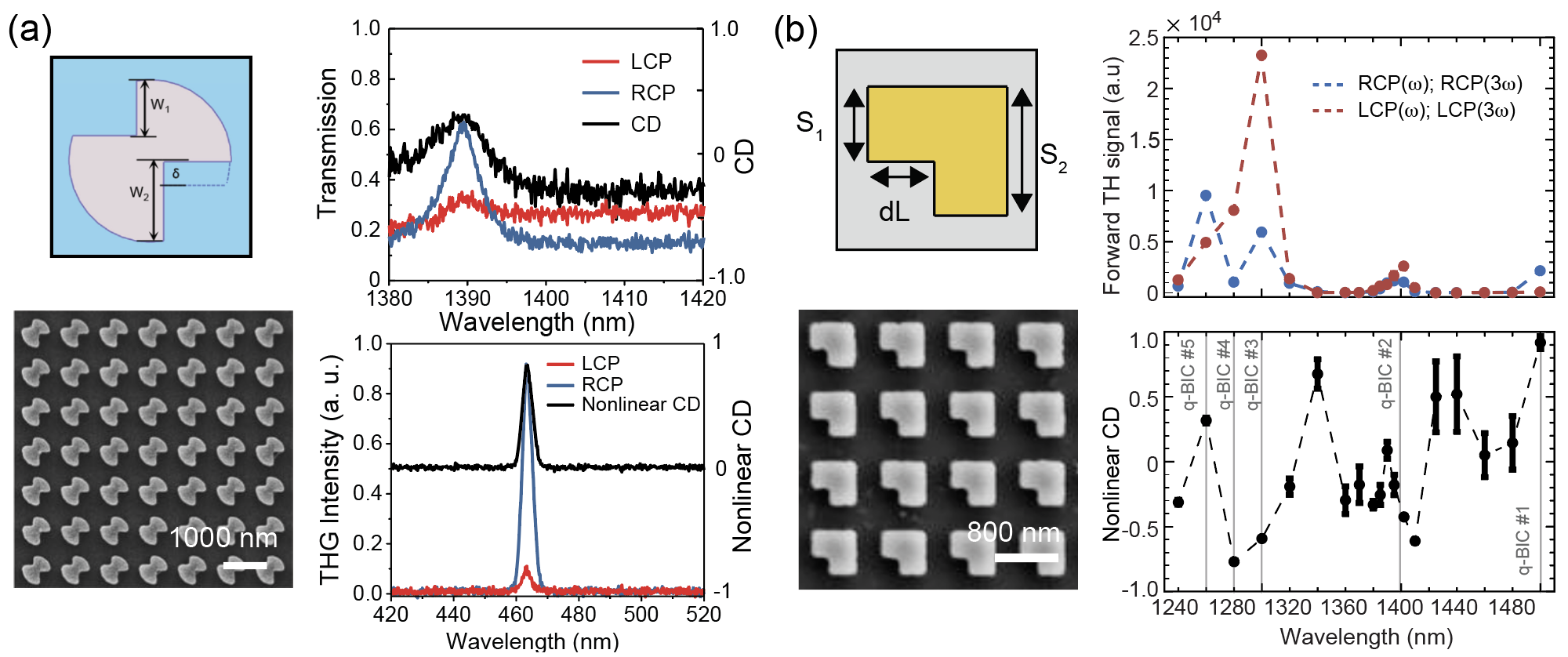}
    \caption{{\bf Nonlinear chiral high-Q dielectric metasurfaces supporting quasi-BICs.} (a) Chiral broken-symmetry nonlinear metasurface hosting a quasi-BIC mode~\cite{shi2022planar}. Left: schematic and SEM images; right: linear transmission and CD (upper) and TH intensity and TH CD (lower) for the LCP and RCP pump in the vicinity of the quasi-BIC resonance. (b) Asymmetric Si metasurface demonstrating record-high TH CD with strong THG efficiency via excitation of quasi-BICs and GMR mode~\cite{koshelev2023resonant}. Left: schematic and SEM images; right: measured forward TH signal and TH CD for the optimized sample with $dL=300$~nm.}
    \label{fig5}
\end{figure}

The first theoretical concept of applying quasi-BICs in dielectric metasurfaces for enhancement of nonlinear CD was  proposed by Gandolfi {\it et al.}~\cite{gandolfi2021near} who considered a Si metasurface composed of asymmetric dimers and placed on a glass substrate hosting quasi-BIC in the near-IR frequency range. The authors demonstrated numerically that such structures can achieve near-unity TH CD in the vicinity of the quasi-BIC mode together with the enhancement of the overall TH signal. This concept was later applied to double-layer photonic crystal slabs with asymmetric dimer-shaped holes in Si film~\cite{liu2022dual}.

The pioneering experimental observations of chiral nonlinear BICs for strong nonlinear CD have been reported only recently~\cite{shi2022planar,koshelev2023resonant}. Shi {\it et al.}~\cite{shi2022planar} studied a Si metasurface composed of a square lattice of asymmetric meta-atoms, with the schematic and SEM images shown in the top and bottom left panels of Fig.~\ref{fig5}a. The in-plane asymmetry of the unit cell due to the parameter $\delta = W_2-W_1$ induced transition from symmetry-protected BIC to quasi-BICs, which were observed in the near-IR range for both LCP and RCP polarizations, see the top right panel of Fig.~\ref{fig5}a. The linear CD reached the value of $0.65$ at the resonance. The nonlinear measurements revealed that the TH intensity is increased dramatically at the quasi-BIC wavelength of $1390$~nm for the RCP signal, contributing to a large TH CD of $0.93$, as can be seen in the bottom right panel of Fig.~\ref{fig5}a. 

Another paper~\cite{koshelev2023resonant} focused on the study of the impact of different kinds of high-$Q$ optical resonances in dielectric metasurfaces on nonlinear optical activity, and maximization of the TH efficiency and TH CD simultaneously by virtue of quasi-BICs and other photonic resonances. The authors studied a Si metasurface composed of L-shaped asymmetric particles with the asymmetry defined as $dL=S_2-S_1$, see the schematic and SEM image in the left panels of Fig.~\ref{fig5}b. The authors demonstrated that the metasurface hosts five quasi-BIC and one GMR mode in the range from $1250$ to $1500$~nm, which the $Q$ factor and wavelength can be gradually shifted by changing the asymmetry parameter $dL$ from $100$ to $300$~nm. The TH signal was maximized for specific values of $dL$ in this range, different for each mode, which was explained by the critical coupling effect due to parasitic surface scattering losses, recently outlined for nonlinear quasi-BIC metasurfaces~\cite{koshelev2019nonlinear,bernhardt2020quasi}. It was shown that the TH conversion efficiency can be maximized for both quasi-BICs and GMR mode reaching the same magnitude, which was attributed to the moderate fabrication quality of the metasurface. The co-polarized TH signals for the LCP and RCP excitation, respectively, measured for the optimized sample satisfying the critical coupling condition were shown to differ substantially in the vicinity of the BIC resonances, as shown in the top right panel of Fig.~\ref{fig5}b. It was observed experimentally that TH CD can be changed gradually by changing the asymmetry parameter $dL$ from large negative to large positive values with keeping the TH intensity large. The maximal measured nonlinear TH CD reached the values of $-0.771$ and $0.918$, which is the record-high demonstrated value for chiral dielectric metasurfaces to the best of our knowledge. We notice that the high impact of nanofabrication defects on the quality of quasi-BIC resonances in dielectric metasurfaces was outlined in the recent work~\cite{kuhne2021fabrication} and just very recently extended to the case of chiral quasi-BIC metasurfaces~\cite{fagiani2023modelling}. 



\section{Conclusion and Outlook}

In the recent years, we observe a rapid growth of the studied in chiral nonlinear metaphotonics due to refocusing of nanophotonics research from plasmonic to dielectric structures hosting versatile Mie and lattice resonances. We believe that several directions will attract attention in this field in the near future. Below we mention briefly only some of those developments including chiral generation of higher-order optical harmonics, chiral quantum optics, chiral polaritonics, chiral cavities, and topological photonics.

High-harmonic generation (HHG) from nanostructured solid materials was observed for the first time only a few years ago~\cite{vampa2017plasmon}, which led to an increased attention to the generation of ultrafast pulses with resonant nanostructured made of dielectric and semiconductor materials, ranging from Si~\cite{zograf2022high} and GaP~\cite{shcherbakov2021generation} to halide perovskites~\cite{tonkaev2023observation}. We anticipate that the studies of chirality combined with HHG will be the next step in this direction, following the earlier observation of bright, phase-matched, extreme ultraviolet circularly-polarized high-harmonics source in gases~\cite{kfir2015generation}. Bridging high nonlinear CD with efficient HHG in nanophotonics could allow for extra control of ultra-fast light-matter interaction in the extreme UV and soft X-ray frequency ranges. The recent work presented a theory of optical chirality for HHG in bulk nonlinear media uncovering the role of noninstantaneous polarization-associated chirality, and time-scale-weighted polarization-associated chirality, and showing that these quantities link the chirality of multichromatic pumps and their generated attosecond pulses~\cite{neufeld2018optical}.


Control of non-classical light and enhancement of quantum nonlinear processes such as spontaneous parametric down-conversion (SPDC) with structured metasurfaces is a hot topic recently emerged in metaphotonics~\cite{solntsev2021metasurfaces}. The current active developments suggest new methods of control of quantum light including measuring photon statistics, quantum entanglement, and detecting single photons. Several very recent studies explored the chiral properties of photons as a new avenue of quantum optics and photonics. Last year, dielectric metasurfaces pumped with circularly-polarized light were used for characterization of the orbital angular momentum of quantum states in the SPDC process~\cite{wang2023characterization}. Another very recent study demonstrated optical spin-orbit interaction in the SPDC process realized through a nonlinear crystal with threefold rotational symmetry.~\cite{wu2023optical} Following these studies, we foresee a variety of theoretical and experimental studies of chirality in quantum light interaction with metastructures empowered by optical resonances.

One of the very latest discoveries in chiral metaphotonics is the field of chiral polaritonics, describing formation of polaritons exhibiting chiral features via interaction between chiral quantum emitters and modes of chiral cavities~\cite{baranov2023toward}. This field is in its infancy, and the very first models have been developed to describe the behavior of chiral polaritons in the formalism of cavity quantum electrodynamics~\cite{sch?fer2023chiral}. We believe that a variety of novel effects will be discovered soon in nonlinear chiral polaritonics that could be enabled by strong coupling between chiral light and nonlinear matter.

Using chiral metasurfaces, one can build very different optical cavities
depending on the desired functionality. More specifically, different pairs of metasurfaces can constitute {\it chiral metacavities} supporting
counter-propagating waves of the same circular polarization. Such cavities based on chiral mirrors, posses remarkable excitation selectivity~\cite{cavity}, for example the intensity of helical standing wave excited by incident LCP waves is by more than two orders of magnitude higher than when being excited by RCP waves. Reciprocally, the Purcell
factor of a right-handed chiral emitter inside such cavity exceeds by more than an order of magnitude that of a left-handed emitter. Recently, a particular example of such metacavity based on aluminum mirror and a metasurfaces consisting of square lattice of silicon triangular prisms was used to realize chiral electroluminescence from a plain perovskite
layer~\cite{cavity2}.

Finally, we wish to mention the emerging field of nonlinear topological photonics that merges the physics of topological phases with nonlinear optics~\cite{apr_topology}.  In topological systems, topological protection in the linear regime is associated with the topological edge states that occur in counter-propagating pairs with opposite spin. Nonlinearity may introduce asymmetry resulting in optical nonreciprocity, such as preferential lasing of a single-mode handedness or chirality,

\section*{Acknowledgements}

Kirill Koshelev and Pavel Tonkaev contributed equally to this work. Yuri Kivshar is indebted to Kristina Frizyuk, Martti Kauranen, Kuniaki Konishi, and Ventsislav Valev for many useful comments and suggestions. This work was supported by the Australian Research Council (grants DP200101168 and DP210101292) and the International Technology Center Indo-Pacific (ITC IPAC) via Army Research Office (contract FA520921P0034). 

\bibliography{refs}

\end{document}